\documentclass[pra,aps,showpacs,twocolumn,twoside]{revtex4}

\usepackage{amsfonts}
\usepackage{amsmath}
\usepackage{amssymb}
\usepackage{graphicx}
\usepackage{psfrag}
\usepackage{color}

\definecolor{labelkey}{gray}{.1}
\definecolor{refkey}{gray}{.1}
\definecolor{dgreen}{rgb}{0.000,0.500,0.000}
\definecolor{purple}{rgb}{0.500,0.000,0.500}
\newcommand{\C}{{\mathbb C}}
\newcommand{\qed}{\hfill\(\square\)}
\newtheorem{theorem}{Theorem}
\newtheorem{lemma}{Lemma}

\begin{document}
\title{Multipartite entanglement in $\bf 2\times 2\times n$ quantum systems}

\author{Akimasa Miyake}
\email{miyake@monet.phys.s.u-tokyo.ac.jp}
\affiliation{\mbox{Department of Physics, Graduate School of Science,
University of Tokyo, Hongo 7-3-1, Bunkyo-ku, Tokyo 113-0033, Japan} \\
\mbox{Quantum Computation and Information Project, ERATO, JST,
Hongo 5-28-3, Bunkyo-ku, Tokyo 113-0033, Japan}}
\author{Frank Verstraete}
\email{frank.verstraete@mpq.mpg.de}
\affiliation{Max-Planck Institut f\"{u}r Quantenoptik,  
Hans-Kopfermann Str. 1, Garching, D-85748, Germany}
\date{July 9, 2003}

%
%
\begin{abstract}
We classify multipartite entangled states in the Hilbert space
\(\mathcal{H}=\C^2 \otimes \C^2 \otimes \C^n \;(n \geq 4)\),
for example the 4-qubit system distributed over 3 parties, under
local filtering operations. We show that there exist nine
essentially different classes of states, giving rise to a
five-graded partially ordered structure, including  the celebrated
Greenberger-Horne-Zeilinger (GHZ) and W classes of three qubits. 
In particular, all $2\times2\times n$-states can be deterministically 
prepared from one maximally entangled state, and some applications 
like entanglement swapping are discussed.
\end{abstract} 

\maketitle

%
\section{Introduction}
\label{sec:intro} Entanglement is the key ingredient of all
applications in the field of quantum information. Due to the
non-local character of the correlations that entanglement induces,
it is expected that entanglement is especially valuable in the
context of many parties. Despite a lot of efforts however, it has
been proven exceedingly hard to get insight into the structure of
multipartite entanglement. Still, the motivation of our work is as
follows. In the bipartite (pure) setting, the entanglement present
in a Bell-Einstein-Podolsky-Rosen (Bell-EPR) state is essentially unique; 
i.e., we can evaluate any bipartite entangled state by the number of 
equivalent Bell pairs, in either a qubit- or a qudit- system, both 
in the single-copy and multiple-copies case.

The situation is totally different in the multipartite setting
however, where interconvertibility under local operations and
classical communication (LOCC) is not expected to hold
\cite{bennett+00}. Multipartite entanglement exhibits a much
richer structure than bipartite entanglement. The first celebrated
example thereof was the 3-qubit GHZ state, called after
Greenberger, Horne and Zeilinger \cite{GHZ89}. This state was
introduced because it allows to disprove the Einstein locality for
quantum systems without invoking statistical arguments such as
needed in the arguments of Bell. Another interesting aspect of
multipartite entanglement was discovered by Wootters {\it et al.}
\cite{coffman+00}. They showed that a quantum state has only a
limited shareability for quantum correlations: the more
bipartite correlations in a state, the less genuine multipartite
entanglement that can be present in the system. This led to the
introduction of the so-called 3-qubit W state \cite{dur+00},
which was shown to be essentially different from the GHZ state as
they are not interconvertible under LOCC even probabilistically.

In this paper, we will generalize these results and present one of
the very few exact and complete results about multipartite quantum
systems, by classifying multipartite entanglement in the 
\(2 \times 2 \times n\) cases. 
Since this include the 4-qubit system distributed over 
{\em 3 parties}, which is the case in e.g., entanglement swapping, 
our results will clarify what kinds of essentially different
multipartite entanglement there exist in this situation, and give 
better understanding for multi-party LOCC protocols.
More specifically, we will address the stochastic LOCC (SLOCC) 
classification of entanglement
\cite{bennett+00,dur+00,verstraete+01_normal,verstraete+01_lsvd,
verstraete+02_3bit,verstraete+02_4bit,miyake03,klyachko02,luque+03_4bit,
luque+03_3trit,jaeger+03}, which is a {\it coarse-grained}
classification under LOCC. Let us consider the single copy of a
multipartite pure state \(|\Psi\rangle\) on the Hilbert space
\({\mathcal H} = \C^{k_1}\otimes \cdots \otimes \C^{k_l}\)
(precisely, in abuse of the notation, we would denote a ray on its
complex projective space \(\C P^{k_{1} \times\cdots\times
k_{l}-1}\) by \(|\Psi\rangle\)),
\begin{equation}
\label{eq:psi}
|\Psi\rangle =\sum_{i_1,\ldots,i_l =0}^{k_1 -1,\ldots,k_l-1}
\psi_{i_1 \ldots i_l} |i_1\rangle \otimes\cdots\otimes |i_l\rangle,
\end{equation}
where a set of \(|i_1\rangle \otimes\cdots\otimes |i_l\rangle\)
constitutes the standard computational basis and it often will be
abbreviated to \(|i_1 \cdots i_l\rangle\). In LOCC, we recognize
two states \(|\Psi\rangle\) and \(|\Psi'\rangle\) which are
interconvertible deterministically, e.g., by local unitary
operations, as equivalent entangled states. On the other hand in
SLOCC, we identify two states \(|\Psi\rangle\) and
\(|\Psi'\rangle\) as equivalent if they are interconvertible
probabilistically, i.e., with a nonvanishing probability, since
they are supposed to be able to perform the same tasks in quantum
information processing but with different success probabilities.
Mathematically, \(|\Psi\rangle\) and \(|\Psi'\rangle\) belong to
the same SLOCC entangled class if and only if they can be
converted to each other by {\it invertible} SLOCC operations,
\begin{equation}
\label{eq:slocc}
|\Psi'\rangle = M_1 \otimes \cdots \otimes M_l |\Psi\rangle,
\end{equation}
where \(M_i\) is any local operation having a nonzero determinant
on the \(i\)-th party \cite{dur+00}, i.e., \(M_i\) is an element
of the general linear group \(GL(k_i,\C)\) (we do not care about
the overall normalization and phase so that we can take its
determinant 1, i.e., \(M_i \in SL(k_i,\C)\).). It can be also said
that an invertible SLOCC operation is a completely positive map
followed by the postselection of one successful outcome.
Mathematically, the SLOCC classification is equivalent to the
classification of orbits generated by a direct product of special
linear groups \(SL(k_1,\C) \times\cdots\times SL(k_l,\C)\). Note
that in the bipartite \(l=2\) case, the SLOCC classification means
the classification just by the Schmidt rank (or equivalently, the
rank of a coefficient "matrix" \(\psi_{i_1 i_2}\) in
Eq.~(\ref{eq:psi})). We will also address the
question of {\it noninvertible} SLOCC operations
(at least one of the ranks of \(M_i\) in Eq.~(\ref{eq:slocc}) is
not full). The set of invertible and noninvertible SLOCC operations
are also called local filtering operations.
Consider the bipartite case as an example: SLOCC entangled classes are
found to be totally ordered in such a way that an entangled class
of the larger Schmidt rank is more entangled than that of the
smaller one, because the Schmidt rank is always decreasing under
noninvertible local operations.

%
%
The paper is organized as follows.
In Sec.~\ref{sec:class}, we classify multipartite \(2 \times 2 \times n\)
pure states under SLOCC, so as to show that nine entangled
classes are hierarchized in a five-graded partial order.
We discuss the characteristics of multipartite entanglement in our
situation in Sec.~\ref{sec:chara}, and extend the classification of
multipartite pure states to mixed states in Sec.~\ref{sec:mixed}.
The conclusion is given in Sec.~\ref{sec:conclude}.

%
\section{Classification of multipartite entanglement}
\label{sec:class}
In this section, we give the complete SLOCC classification of
multipartite entanglement in \(2 \times 2 \times n\) cases.
Moreover, we present a convenient criterion to distinguish
inequivalent entangled classes by SLOCC invariants.

%
%
\subsection{Five-graded partial order of nine entangled classes}
\label{sec:theorem}
We show that there are nine entangled classes and they constitute
five-graded partially ordered structure under noninvertible SLOCC operations.

\begin{figure}[t]
\begin{center}
\includegraphics[clip]{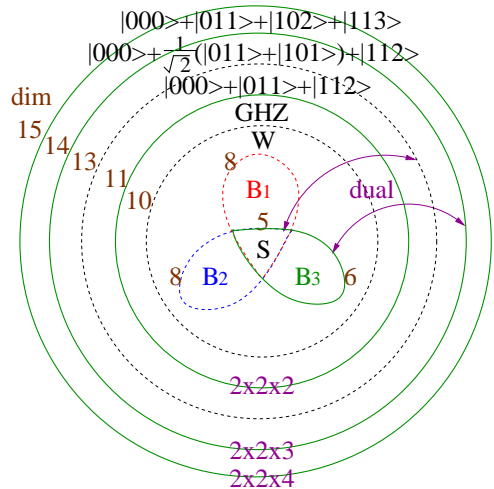}
\caption{ The onion-like classification of multipartite entangled classes
(SLOCC orbits) in the Hilbert space \(\mathcal{H}=\C^2 \otimes \C^2 \otimes
\C^n \;(n \geq 4)\).
There are nine classes divided by "onion skins" (the orbit closures).
The pictures for \(2 \times 2 \times n (n > 4)\) cases are essentially same
although the dimensions of SLOCC orbits are different.
These classes merge into four classes, divided by the skins of the solid line,
in the "bipartite" (AB)-C picture.
Note that although noninvertible SLOCC operations generally cause
the conversions inside the onion structure, an outer class can not
necessarily convert into its neighboring inner class (cf.
Fig.~\ref{fig:part}).}
\label{fig:onion}
\end{center}
\end{figure}

\begin{figure}[t]
\begin{center}
\includegraphics[clip]{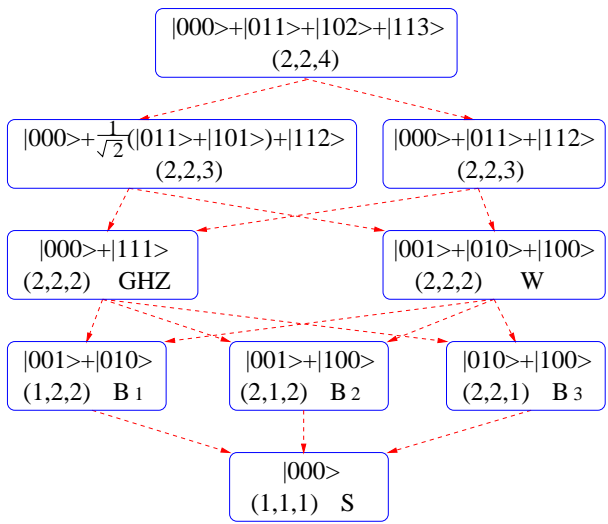}
\caption{
The five-graded partially ordered structure of nine entangled classes
in the \(2 \times 2 \times n\) (\(n \geq 4\)) case.
Every class is labeled by its representative, its set of local ranks, and
its name.
Noninvertible SLOCC operations, indicated by dashed arrows, degrade
higher entangled classes into lower entangled ones.}
\label{fig:part}
\end{center}
\end{figure}

\begin{theorem}
\label{th:main} Consider pure states in the Hilbert space
\({\mathcal H}= \C^2 \otimes \C^2 \otimes \C^n \;(n \geq 4)\),
they are divided into nine entangled classes, seen in
Fig.~\ref{fig:onion}, under invertible SLOCC operations. These
nine entangled classes constitute the five-graded partially
ordered structure of Fig.~\ref{fig:part}, where noninvertible
SLOCC operations degrade higher entangled classes into lower
entangled ones.
\end{theorem}


Some remarks are given before its proof. The theorem gives the
complete classification of multipartite pure entangled states in
\(2 \times 2 \times n \;(n \geq 4)\) cases. It naturally contains
the classification for the \(2 \times 2 \times 2\) (3-qubit) case
\cite{dur+00,verstraete+02_3bit,miyake03} and the \(2 \times 2
\times 3\) case \cite{miyake03}. We find that SLOCC orbits are
added outside the onion-like picture (Fig.~\ref{fig:onion}) and
the partially ordered structure (Fig.~\ref{fig:part}) becomes
higher, as the third party Clare has her larger subsystem.
Remarkably, for the \(2 \times 2 \times n \;(n \geq 4)\) cases,
the generic class is one "maximally entangled" class located on
the top of the hierarchy. This is a clear contrast with the
situation of the \(2 \times 2 \times 2\) and \(2 \times 2 \times
3\) cases, where there are two different entangled classes on its
top. It suggests that, even in the multipartite situation, there
is a unique entangled class which can serve as resources to create
any entangled state, if the Clare's subsystem is large enough.
This will be proven in Sec.~\ref{sec:chara}.

We note that it is sufficient to consider the \(2 \times 2 \times
4\) case in the proof of the theorem, since Clare can only have
support on a 4-dimensional subspace. This is an analogy with the
bipartite \(k \times k'\) (\(k < k'\)) case whose SLOCC
classification is equivalent to that of the \(k \times k\) case,
because the SLOCC-invariant Schmidt local rank takes at most
\(k\). In any \(2\times 2\times n \;(n \geq 4)\) case, the
partially ordered structure of multipartite entanglement consists
of nine finite classes. Our result not only describes the
situation that only Clare has the abundant resources, but also
would be useful in analyzing entanglement of two-qubit mixed
states attached with an environment (the rest of the world), which
could e.g. be used to  analyze the power of an eavesdropper in
quantum cryptography.

Let us consider the situation where Alice and Bob
are considered as one party (or, one of Alice and Bob comes to have two qubits)
and call it the "bipartite" (AB)-C picture.
When two parties have two qubits for each, the onion-like structure of
Fig.~\ref{fig:onion} becomes coarser.
The nine entangled classes merge into four classes, and the structure
coincides with that of the bipartite \(4 \times 4\) case.
We see that we can perform LOCC operations more freely in the bipartite
situation.
Likewise, in the bipartite A-(BC) or B-(AC) pictures, the onion-like structure
coincides with that of the \(2 \times 8\) (i.e., \(2 \times 2\)) case
so that just two entangled classes, divided by the onion skin of \(B_1\) or
\(B_2\) respectively, remain.

On the other hand, it can be said that the SLOCC-invariant onion structure
of the \(2 \times 2\times 4\) case is a coarse-grained one of the
4 qubits (\(2\times 2\times 2\times 2\)) case (see also
Ref.\cite{verstraete+02_4bit,luque+03_4bit,note_4bit}), i.e., the former is
embedded into the latter in the same way as the structure of the bipartite
\(4 \times 4\) case is embedded into that of the \(2 \times 2 \times 4\) case.
So, if two 4-qubit states belong to different classes
in the \(2 \times 2 \times 4\) classification, these states must be also
different in the 4-qubit classification.
It would be interesting to note that the 4-qubit entangled states are divided
into infinitely many classes \cite{dur+00,verstraete+02_4bit,miyake03}, in
comparison with finitely many classes of the \(2 \times 2\times 4\) case.
In other words, there are infinitely many orbits in the 4 qubits case
between some onion skins, while there exists one orbit in the
\(2 \times 2 \times 4\) case.
This suggests that a drastic change occurs in the structure of multipartite
entanglement even when a party comes to have two qubits in hands
\cite{note_change}.

%
%

Now, we give the proof of the Theorem~\ref{th:main} in two different,
algebraic (in Sec.~\ref{sec:theorem}) and geometric (in
Sec.~\ref{sec:geometry}), ways.
Readers who are interested just in applying our results can skip to
Sec.~\ref{sec:criterion}, where a convenient criterion for distinguishing
nine classes is given.

{\em Proof.  }
We first give an algebraic proof, utilizing the matrix analysis
(cf. Ref.~\cite{verstraete+01_normal,verstraete+01_lsvd,verstraete+02_3bit,
verstraete+02_4bit}).
Any state is parameterized by a three index tensor $\psi_{i_1 i_2 i_3}$
with $i_1,i_2\in\{0,1\}$ and $i_3\in\{0,1,2,3\}$. This tensor can be
rewritten as a $4\times 4$ matrix $\tilde{\Psi}=(\psi_{(i_1 i_2) i_3})$
by concatenating the indices $(i_1,i_2)$. Next we define the matrix \(R\) as
\begin{equation}
\label{eq:R}
R=T\tilde{\Psi},
\end{equation}
where $T$ is defined as
\begin{equation}
\label{eq:T}
T=\frac{1}{\sqrt{2}}
\left(\begin{array}{cccc}
1&0&0&1\\ 0&i&i&0\\ 0&-1&1&0\\ i&0&0&-i
\end{array}\right).
\end{equation}
Let us observe that both \(2 \times 2\) matrices $M_1$ and $M_2$ belong to
$SL(2,\C)$ if and only if $O = T(M_1\otimes M_2)T^\dagger\in SO(4,\C)$ and
$\det(M_1)=\det(M_2)=1$, because of a consequence of an accident in the
Lie group theory: $SL(2,\C)\otimes SL(2,\C)\simeq SO(4,\C)$
(cf. $SU(2)\otimes SU(2)\simeq SO(4)$ ).
Accordingly, we see that a SLOCC transformation of Eq.~(\ref{eq:slocc})
results in a transformation
\begin{equation}
R'=O R M_3^T.
\end{equation}
Thus, our problem is equivalent to finding appropriate normal forms for
the complex $4\times 4$ matrix $R$ under left multiplication with a complex
orthogonal matrix $O \in SO(4,\C)$ and right multiplication with an arbitrary
\(4\times 4\) matrix $M_3^T \in SL(4,\C)$.

If the matrix $R$ has full rank, it is enough to operate $M_3$ chosen to be
$T^\dagger (R^{-1})^T$. As a result, the state \(\tilde{\Psi}\) is
(proportional to) the identity matrix \(\openone\), or
\begin{equation}
|000\rangle+|011\rangle+|102\rangle+|113\rangle,
\end{equation}
the representative of the highest class in the hierarchy.

Suppose however that the rank of $R$ is three. As a first step,
$R$ can always be multiplied left by a permutation matrix
and right by $M_3^T$ so as to yield an $R$ of the form
\begin{equation}
R=\left(\begin{array}{cccc}
1&0&0&0\\ 0&1&0&0\\ 0&0&1&0\\ \alpha&\beta&\gamma&0
\end{array}\right).
\end{equation}
Suppose $\alpha\neq \pm i$, then it can easily be checked that
left multiplication by the complex orthogonal matrix
\begin{equation}
O = \left(\begin{array}{cccc}
1/\sqrt{\alpha^2 +1}&0&0&\alpha/\sqrt{\alpha^2 +1}\\ 0&1&0&0\\
0&0&1&0\\ -\alpha/\sqrt{\alpha^2 +1}&0&0&1/\sqrt{\alpha^2 +1}
\end{array}\right)
\end{equation}
and right multiplication with
\begin{equation}
M_3^T = \left(\begin{array}{cccc}
1&-\alpha\beta/(\alpha^2+1)&-\alpha\gamma/(\alpha^2+1)&0\\
0&1&0&0\\ 0&0&1&0\\ 0&0&0&1
\end{array}\right)
\end{equation}
yield a new $R$ of the form
\begin{equation}
R=\left(\begin{array}{cccc}
1&0&0&0\\ 0&1&0&0\\ 0&0&1&0\\ 0&\beta'&\gamma'&0
\end{array}\right).
\end{equation}
Exactly the same can be done in the case where $\beta,\gamma \neq\pm i$, and
therefore we only have to consider the case where
$\alpha,\beta,\gamma \in\{0,i,-i\}$. It can however be checked that in the case
that when 2 or 3 elements $\alpha,\beta,\gamma$ are not equal to zero,
a new $R$ can be made where all $\alpha,\beta,\gamma$ become equal to zero:
this can be done by first multiplying $R$ with orthogonal matrices of the kind
\begin{equation}
O=\left(\begin{array}{cccc}
1&0&0&0\\ 0&1&0&0\\ 0&0&1/\sqrt{2}&- 1/\sqrt{2}\\ 0&0&1/\sqrt{2}&1/\sqrt{2}
\end{array}\right),
\end{equation}
and repeating the procedure outlined above. There remains the case
where exactly one of the elements is equal to $\pm i$. Without
loss of generality, we assume that $(\alpha,\beta,\gamma)=(i,0,0)$
(this is possible because one can do permutations (with signs) by
appropriate $O\in SO(4)$ and $M_3$). This case is fundamentally
different from the one where all $\alpha,\beta,\gamma$ are equal to zero
as the corresponding matrix $R^TR$ has rank $2$ as opposed to rank $3$ of
$R$. There is no way in which this behavior can be changed by
multiplying $R$ left and right with appropriate transformations,
and we therefore have identified a second class (which is clearly
of measure zero: a generic rank 3 state $R$ will also yield a rank
3 $R^TR$).

It is now straightforward to construct a representative state of
each class. As a representative of the major class in the rank 3 \(R\),
we choose the state
\begin{equation}
|000\rangle+\frac{1}{\sqrt{2}}(|011\rangle+|101\rangle)+|112\rangle.
\end{equation}
As a representative of the minor class in the rank 3 \(R\), we choose
the state
\begin{equation}
|000\rangle+|011\rangle+|112\rangle,
\end{equation}
as it makes clear that the states in this class can be transformed to have
3 terms in some product basis (as opposed to the states in the major class
that can be transformed to have 4 product terms).

The case where $R$ has rank $2$ can be solved in a completely
analogous way. Exactly the same reasoning leads to the following
four possible normal forms for $R$:
\begin{equation}
\begin{split}
&\left(\begin{array}{cccc}
1&0&0&0\\ 0&1&0&0\\ 0&0&0&0\\ 0&0&0&0
\end{array}\right), \quad
\left(\begin{array}{cccc}
1&0&0&0\\ 0&1&0&0\\ 0&0&0&0\\ i&0&0&0
\end{array}\right), \\
&\left(\begin{array}{cccc}
1&0&0&0\\ 0&1&0&0\\ 0&i&0&0\\ i&0&0&0
\end{array}\right), \quad
\left(\begin{array}{cccc}
1&0&0&0\\ 0&1&0&0\\ 0&-i&0&0\\ i&0&0&0
\end{array}\right).
\end{split}
\end{equation}
Note that the last two cases cannot be transformed into each other due to the
constraint that $O$ has determinant $+1$. The corresponding
representative states are easily obtained by choosing symmetric ones:
\begin{align}
&|000\rangle+|111\rangle, \\
&|001\rangle+|010\rangle+|100\rangle, \\
&|000\rangle+|011\rangle, \\
&|000\rangle+|101\rangle.
\end{align}
The first state is the celebrated Greenberger-Horne-Zeilinger (GHZ) state,
the second one the W state named in \cite{dur+00} for the \(3\)-qubit case,
and the remaining ones represent biseparable \(B_i \;(i=1,2)\) states with
only bipartite entanglement between Bob and Clare, or Alice and Clare,
respectively.

As a last class, we have to consider the one where $R$ has rank
equal to $1$. This leads to the following two possible normal forms for $R$:
\begin{equation}
\left(\begin{array}{cccc}
1&0&0&0\\ 0&0&0&0\\ 0&0&0&0\\ 0&0&0&0
\end{array}\right), \quad
\left(\begin{array}{cccc}
1&0&0&0\\0&0&0&0\\0&0&0&0\\i&0&0&0
\end{array}\right).
\end{equation}
The corresponding states are given by
\begin{align}
&|000\rangle+|110\rangle, \\
&|000\rangle,
\end{align}
which are the biseparable \(B_3\) state and the completely separable \(S\)
state, respectively.
This ends the complete classification.

It remains to be proven that any state that is higher in the
hierarchy of Fig.~\ref{fig:part} can be transformed to all the other
ones that are strictly lower. The first step downwards is evident from the
fact that right multiplication of a rank 4 $R$ with a rank
deficient $M_3$ can yield whatever $R$ of rank $3$. In going from a rank $3$
$R$ of the major class to a rank $2$, the state
$|000\rangle+(|011\rangle+|101\rangle)/\sqrt{2}+|112\rangle$ can
be transformed into the GHZ state by a projection of Clare on the subspace
$\{|0\rangle,|2\rangle\}$ and into the
W state by Clare implementing the POVM element
\begin{equation}
\left(\begin{array}{cccc}
1&0&0&0\\ 0&1&0&0\\ 0&i&0&0\\ 0&0&0&0
\end{array}\right).
\end{equation}
From a rank 3 \(R\) of the minor class, the GHZ state can easily
be constructed by a projection of Clare on her
$\{|1\rangle,|2\rangle\}$ subspace, while the W state is obtained
by Clare projecting on her $\{|0\rangle,|1\rangle+|2\rangle\}$
subspace. Finally, the conversion of the GHZ and W states to the
Bell state among two parties (the biseparable state), as well as
that of the Bell state to the completely separable state, is
straightforward. \qed

The proof not only gives a constructive transformations to
representatives of nine entangled classes, but also suggests a
very simple way of determining to which class a given state
belongs. One has to calculate the rank $r(.)$ of the matrices $R$
(see Eq. (\ref{eq:R})), of $R^T R$, and of the reduced density
matrix $\rho_1$. One gets the following classification:
\begin{equation}
\label{table}
\begin{array}{|c|c|c|c|}
\hline {\rm Class} & r(R) & r(R^TR) & r(\rho_1)\\
\hline |000\rangle+|011\rangle+|102\rangle+|113\rangle & 4 & 4 & 2\\
\hline
|000\rangle+\frac{1}{\sqrt{2}}(|011\rangle+|101\rangle)+|112\rangle
& 3 & 3 & 2\\ \hline |000\rangle+|011\rangle+|112\rangle & 3 & 2 & 2\\
\hline |000\rangle+|111\rangle & 2 & 2 & 2\\ \hline
|001\rangle+|010\rangle+|100\rangle & 2 & 1 & 2\\ \hline
|000\rangle+|101\rangle & 2 & 0 & 2\\ \hline
|000\rangle+|011\rangle & 2 & 0 & 1\\ \hline
|000\rangle+|110\rangle & 1 & 1 & 2\\ \hline |000\rangle & 1 & 0 & 1\\
\hline
\end{array}
\end{equation}

Note that the representative states in the GHZ-type classes were
chosen to be the ones with {\em maximal} entanglement: following
\cite{verstraete+01_normal}, the states with maximal entanglement
in a SLOCC class are the ones for which all local density
operators are proportional to the maximally mixed state. This is
in accordance with the intuition that the local disorder or
entropy is proportional to the entanglement present in the (pure)
state.

\subsection{Geometry of nine entangled classes}
\label{sec:geometry}
We explore how the whole Hilbert space is geometrically divided into
different nine classes, drawn in the onion-like picture Fig.~\ref{fig:onion}.
This subsection can be seen as an alternative proof of the theorem
in Sec.~\ref{sec:theorem} by a geometric way.

We utilize a duality between the set of separable states and the set of
entangled states in order to classify multipartite entangled states under
SLOCC \cite{miyake03}.
The set \(S\) of completely separable states is the smallest closed subset,
as seen in Fig.~\ref{fig:onion}.
In many cases (such as the \(l\)-qubit cases) of interest to quantum
information, its dual set is the largest closed subset which consists of all
degenerate entangled states, and is given by the zero hyperdeterminant
\({\rm Det}\Psi =0\).
We readily see that, in the bipartite \(k \times k\) case, the set
\(S\) is the smallest subset of the Schmidt rank 1, while its dual set is
the largest subset where the Schmidt rank is not full (i.e., \(\det \Psi =0\)).

However, the entangled states in \({\mathcal H}=\C^2 \otimes \C^2 \otimes
\C^n \) (\(n \geq 4\)) have a peculiar structure from a geometric viewpoint.
It is not the case here that the largest subset is dual to the smallest
subset \(S\).
Indeed, the largest subset is dual to (the closure of) the set \(B_3\) of
the biseparable states, i.e.,the second smallest closed subset of dimension 6
in Fig.~\ref{fig:onion}.
The dual set of \(S\) is the second largest subset of dimension 13.
The reason will be explained later.
Significantly, this suggests that for the \(2 \times 2\times n\)
(\(n \geq 4\)) cases, there are no hyperdeterminants in the Gelfand
{\it et al.}'s sense; in other word, the onion structure will not change
any more for \(n \geq 4\).
This is intuitively because the subsystem of one party
is too large, compared with the subsystems of the other parties.
Remember that it is again an analogy to the bipartite \(k \times k'\) case
(\(k < k'\)), where there is no determinant but its onion structure
remains unchanged from that of the \(k \times k\) case.

In general, the hyperdeterminants can be defined for
\({\mathcal H}=\C^{k_1}\otimes \cdots \otimes \C^{k_l}\),
if and only if
\begin{equation}
k_{i} -1 \leq \sum_{j \ne i} (k_{j} -1) \quad \forall i = 1, \ldots, l
\end{equation}
are satisfied \cite{gelfand+94,miyake03}.
Of course, in the bipartite cases, this condition suggests that
the determinants can be defined just for square (\(k_1 = k_2\)) matrices
as usual.
Instead, in the \(2 \times 2 \times 4\) case, the zero locus of the
ordinary determinant of degree 4 for the "flattened" matrix \(\tilde{\Psi}\),
\begin{equation}
\label{eq:det224}
\left| \begin{array}{cccc}
\psi_{000} & \psi_{001} & \psi_{002} & \psi_{003} \\
\psi_{010} & \psi_{011} & \psi_{012} & \psi_{013} \\
\psi_{100} & \psi_{101} & \psi_{102} & \psi_{103} \\
\psi_{110} & \psi_{111} & \psi_{112} & \psi_{113}
\end{array}\right|
\;\left(= \det \tilde{\Psi}\right),
\end{equation}
gives the equation of the largest closed subset.
Note that it is the SLOCC invariant for the bipartite \(4 \times 4\) format
as well as the tripartite \(2 \times 2 \times 4\) format.
It means that the largest subset is dual to the set \(B_3\) of
the biseparable states, i.e., the set of the separable states in the
"bipartite" (AB)-C picture.
We should stress that this duality itself is valid in any
\(2 \times 2 \times n \;(n \geq 4)\) case, regardless of the absence
of the (hyper)determinant.

Next, let us show that the dual set of \(S\) is the second largest subset
for the \(2 \times 2 \times 4\) case.
In order to decide the dual set of \(S\),
we seek for the state \(|\Psi\rangle\) included in the hyperplane (the
orthogonal 1-codimensional subspace) tangent at a completely separable state
\(|x\rangle\) (see Ref.~\cite{miyake03} in detail.).
Mathematically speaking, we should decide the condition for \(|\Psi\rangle\)
such that a set of equations,
\begin{equation}
\label{eq:critical}
\left\{\mbox{  }
\begin{split}
& F(\Psi,x) = \sum_{i_1, i_2, i_3=0}^{1, 1, 3}
\psi_{i_1 i_2 i_3} x^{(1)}_{i_1} x^{(2)}_{i_2} x^{(3)}_{i_3} =  0, \\
& \frac{\partial}{\partial x^{(j)}_{i_j}} F(\Psi,x) = 0
\quad \forall  j, i_j,
\end{split}
\right.
\end{equation}
has at least a nontrivial solution \(x = (x^{(1)},x^{(2)},x^{(3)})\) of
every \(x^{(j)}\ne 0\).
For simplicity, let us suppose that the point of tangency is
the completely separable state \(|000\rangle\) (i.e.,
\(x^{(1)}_{0}=x^{(2)}_{0}=x^{(3)}_{0}=1\), \(\mbox{others}=0\)),
the corresponding state \(|\Psi\rangle\) should satisfy
\begin{equation}
\label{eq:psi_dual}
|\Psi\rangle \in
\{\psi_{000} = \psi_{100} = \psi_{010} = \psi_{001}
= \psi_{002} = \psi_{003} = 0\},
\end{equation}
according to Eq.~(\ref{eq:critical}).
We find that the state \(|\Psi\rangle\) should belong to the class of
dimension 13, because any state,
\begin{align}
|\Psi\rangle = &\psi_{011}|011\rangle +\psi_{012}|012\rangle +
\psi_{013}|013\rangle +\psi_{101}|101\rangle \nonumber\\
&+\psi_{102}|102\rangle +\psi_{103}|103\rangle +\psi_{110}|110\rangle
+\psi_{111}|111\rangle \nonumber\\
&+\psi_{112}|112\rangle +\psi_{113}|113\rangle,
\end{align}
in Eq.~(\ref{eq:psi_dual}) can convert to its representative
\(|011\rangle + |102\rangle + |113\rangle\) under invertible SLOCC operations.

In brief, we find that the 14 dimensional largest subset is the dual set of
the biseparable states \(B_3\), and the 13 dimensional second largest subset
is the dual set of the completely separable states \(S\).
Moreover, we notice that the inside of the largest subset, given by zero
locus of Eq.~(\ref{eq:det224}), is equivalent to the structure of the
\(2 \times 2 \times 3\) case (since the local rank for Clare should be
less than or equal to 3), which has already been clarified in
Ref.~\cite{miyake03}.
That is how we obtain the onion-like picture of Fig.~\ref{fig:onion}.
In general, we can take advantage of all kinds of the dual pairs for sets
(typically, one is a large set and the other is a small set), in order to
distinguish inequivalent entangled classes. This strategy will be explored
elsewhere \cite{miyake+prep}.

%
%
\subsection{Convenient criterion to distinguish nine entangled classes}
\label{sec:criterion}
We give a convenient criterion to distinguish nine entangled classes
by a complete set of SLOCC invariants.
Let us denote local ranks of the reduced density matrices \(\rho_1,\rho_2\),
and \(\rho_3\) such as
\begin{equation}
\rho_i = {\rm tr}_{\forall j \ne i}(|\Psi\rangle\langle\Psi|)
\quad i=1,2,3,
\end{equation}
by the 3-tuples \((r_1,r_2,r_3)\).
These local ranks are always useful SLOCC invariants.
In the bipartite setting, the 2-tuples \((r_1,r_2)\) are enough
to distinguish entangled classes, for both \(r_1\) and \(r_2\) are
indeed nothing but the Schmidt rank.
In the multipartite setting, however, we need more SLOCC invariants in
addition to the set of the local ranks.

The proof of the Theorem~\ref{th:main} in Sec.~\ref{sec:theorem}
has suggested that a complete set of SLOCC invariants is the rank of \(R\)
in Eq.~(\ref{eq:R}) (i.e., \(r_3\)), rank of \(R^{T}R\),
and \(r_1\) (alternatively, \(r_2\)).
Although we have successfully found the rank of \(R^{T} R\) as an additional
SLOCC invariant, this is specific to the substructure associated with 2 qubits,
i.e., to a homomorphism $SL(2,\C)\otimes SL(2,\C)\simeq SO(4,\C)$.

In the following, we introduce another complete set of SLOCC invariants,
since it also gives an insight about how entanglement measures, distinguishing
entangled classes, are derived in general.
The set consists of polynomial invariants (hyperdeterminants
\cite{miyake03,gelfand+94}) adjusted to smaller formats, as well as
3-tuples \((r_1,r_2,r_3)\) of the local ranks.
The criterion reflects the onion structure drawn in Fig.~\ref{fig:onion},
and suggests that we can utilize the results of the SLOCC
classification for smaller formats recursively as if we were skinning
the onion recursively.

%
%
Any pure state in \({\mathcal H}=\C^2 \otimes \C^2 \otimes \C^n\) is written
in the form,
\begin{equation}
|\Psi\rangle = \sum_{i_1,i_2,i_3 =0}^{1,1,n-1} \psi_{i_1 i_2 i_3}
|i_1\rangle\otimes |i_2\rangle\otimes |i_3\rangle.
\end{equation}

First we calculate a set \((r_1,r_2,r_3)\) of the SLOCC-invariant local ranks
of the reduced density matrices.

(i) In the \((2,2,4)\) case, we find that the state \(|\Psi\rangle\) belongs
to the generic class of dimension 15 (the dimension is indicated for
readers' convenience, but it is the one for the \(2\times 2 \times 4\) case.).

(ii) In the \((2,2,3)\) case, there are two possibilities.
Changing the local basis for Clare, we can always choose all new
\(\psi_{i_1 i_2 i_3}=0 \;(i_3 \geq 3)\).
We evaluate the hyperdeterminant of degree 6 for the new,
\(2\times 2\times 3\) formated \(\psi_{i_1 i_2 i_3}\),
\begin{multline}
\label{eq:hdet223}
{\rm Det}\Psi_{2 \times 2 \times 3}=
\left|\begin{array}{ccc}
\psi_{000} & \psi_{001} & \psi_{002} \\ \psi_{010} & \psi_{011} & \psi_{012} \\
\psi_{100} & \psi_{101} & \psi_{102} \\
\end{array}\right|
\left|\begin{array}{ccc}
\psi_{010} & \psi_{011} & \psi_{012} \\ \psi_{100} & \psi_{101} & \psi_{102} \\
\psi_{110} & \psi_{111} & \psi_{112} \\
\end{array}\right| \\
-
\left|\begin{array}{ccc}
\psi_{000} & \psi_{001} & \psi_{002} \\ \psi_{010} & \psi_{011} & \psi_{012} \\
\psi_{110} & \psi_{111} & \psi_{112} \\
\end{array}\right|
\left|\begin{array}{ccc}
\psi_{000} & \psi_{001} & \psi_{002} \\ \psi_{100} & \psi_{101} & \psi_{102} \\
\psi_{110} & \psi_{111} & \psi_{112} \\
\end{array}\right|.
\end{multline}
If \({\rm Det}\Psi_{2\times 2\times 3} \ne 0\), then \(|\Psi\rangle\)
belongs to the major class of dimension 14.
Otherwise (i.e., \({\rm Det}\Psi_{2\times 2\times 3} =0\)), it belongs to
the minor class of dimension 13.

(iii) In the \((2,2,2)\) case, there are also two possibilities.
Changing the local basis for Clare, we can always choose
all new \(\psi_{i_1 i_2 i_3} =0 \; (i_3 \geq 2)\).
We evaluate the hyperdeterminant of degree 4 (its absolute value is
also known as the 3-tangle \cite{coffman+00}) for the \(2\times 2\times 2\)
formated \(\psi_{i_1 i_2 i_3}\),
\begin{align}
\label{eq:hdet222}
&{\rm Det}\Psi_{2 \times 2 \times 2} \nonumber\\
&= \psi_{000}^2 \psi_{111}^2 + \psi_{001}^2 \psi_{110}^2
+ \psi_{010}^2 \psi_{101}^2 + \psi_{100}^2 \psi_{011}^2 \nonumber\\
& -2(\psi_{000}\psi_{001}\psi_{110}\psi_{111}+
\psi_{000}\psi_{010}\psi_{101}\psi_{111}\nonumber\\
& + \psi_{000}\psi_{100}\psi_{011}\psi_{111}+
\psi_{001}\psi_{010}\psi_{101}\psi_{110} \nonumber\\
& + \psi_{001}\psi_{100}\psi_{011}\psi_{110}+
\psi_{010}\psi_{100}\psi_{011}\psi_{101}) \nonumber\\
& + 4 (\psi_{000}\psi_{011}\psi_{101}\psi_{110} +
\psi_{001}\psi_{010}\psi_{100}\psi_{111}).
\end{align}
Likewise, if \({\rm Det}\Psi_{2\times 2\times 2} \ne 0\), then \(|\Psi\rangle\)
belongs to the GHZ class of dimension 11.
Otherwise, it belongs to the W class of dimension 10.

(iv) In the \((1,2,2)\), \((2,1,2)\), and \((2,2,1)\) cases,
\(|\Psi\rangle\) belongs to the biseparable \(B_1\), \(B_2\), and
\(B_3\) class of dimension 8, 8, and 6, respectively.

(v) In the \((1,1,1)\) case, \(|\Psi\rangle\) belongs to the completely
separable class \(S\) of dimension 5.

In this manner, we can immediately check which class a given state
\(|\Psi\rangle\) belongs to.
We remark that the representatives of nine entangled classes
in previous subsections have been chosen with the help of hyperdeterminants;
the "GHZ-like" representatives are chosen to maximize the absolute
value of (hyper)determinants in Eqs.(\ref{eq:det224}), (\ref{eq:hdet223}),
and (\ref{eq:hdet222}), which are entanglement monotones under general LOCC
\cite{verstraete+01_normal,miyake03} (cf. Ref.~\cite{vidal00,barnum+01}).

%
\section{Characteristics of multipartite entanglement}
\label{sec:chara}
%

%
%
\subsection{LOCC protocols as noninvertible flows}
The recent trend of experimental quantum optics reaches the stage
that we can manipulate two Bell states collectively. LOCC
protocols involving local collective operations over two Bell
states are key procedures in , for example, entanglement swapping
\cite{zukowski+93, pan+98} (a building block of quantum
communication protocols like quantum teleportation
\cite{bennett+93} and the quantum repeater \cite{briegel+98}) and
the creation of multipartite GHZ and W states. Although there
appear 4 particles (qubits), these can be seen as LOCC operations
in 3 parties (\({\mathcal H}= \C^2 \otimes \C^2 \otimes \C^4\))
because the third party Clare has initially two particles, each of
which is in a Bell state with another particle on Alice's or Bob's
side respectively, and locally performs collective operations on
them.

Entanglement swapping is the LOCC protocol where the initial state
is prepared as two Bell pairs shared among Alice, Bob, and Clare
in the manner described above. We note that two Bell pairs are
equivalent to the representative of the generic entangled class of
dimension 15,
\begin{align}
\label{eq:2epr} &|\mbox{2 Bell}\rangle = (|00\rangle +
|11\rangle)_{A C_{1}} \otimes
(|00\rangle + |11\rangle)_{B C_{2}} \nonumber\\
&= |00(00)\rangle+|01(01)\rangle+|10(10)\rangle+|11(11)\rangle_{ABC_{12}}.
\end{align}
\(|\mbox{2 Bell}\rangle\) is also equivalent to
\(\sum_{i=0}^{3}|\Phi_i\rangle_{AB}\otimes |\Phi_i\rangle_{C_{12}}\), 
where a set of \(|\Phi_i\rangle \;(i=0,1,2,3)\) is the standard Bell basis. 
So, this protocol can create the biseparable \(B_3\) state which contains 
maximal entanglement (a Bell pair) between Alice and Bob,
\begin{equation}
(|00\rangle + |11\rangle)_{AB} \otimes (|(00)\rangle+|(11)\rangle)_{C_{12}},
\end{equation}
by Clare's local collective Bell measurement (any 
\(|\Phi_i\rangle_{AB}\) corresponding to the outcome \(i\) of her
Bell measurement is equivalent to \((|00\rangle
+|11\rangle)_{AB}\) under LOCC). Thus, entanglement swapping can
be seen as a protocol creating isolated (maximal) entanglement
between Alice and Bob from generic entanglement. In other words,
it is given by a downward flow in Fig.~\ref{fig:part} from the
generic class to the biseparable class \(B_3\). Now, we readily
find that the entanglement swapping protocol is
(probabilistically) successful even when we initially prepare
other 4-qubit entangled states in the generic class.

On the other hand, two Bell pairs can create two different kinds
of genuine 3-qubit entanglement, GHZ and W by Clare's local
collective operations. These LOCC protocols are given by the
downward flow, in Fig.~\ref{fig:part}, from the generic classes to
the GHZ and W class, respectively.

That is how we see that important LOCC protocols in quantum information are
given as noninvertible (downward) flows in the partially ordered structure,
such as Fig.~\ref{fig:part}, of multipartite entangled classes.
So, we expect that the SLOCC classification can give us an insight
in looking for new novel LOCC protocols by means of several entangled
states over multiparties.

\subsection{Two Bell pairs create any state with certainty.}
We show that two Bell pairs are powerful enough to create any
state {\it with certainty} in our \(2 \times 2\times n\) cases. We
find that this is also the case when one of multiparties has a
half of the total Hilbert space.

\begin{theorem}
Consider pure states in the Hilbert space \({\mathcal H}= \C^2
\otimes\C^2 \otimes \C^n\). Two Bell pairs, the representative of
the generic class, can create any state \(|\Psi\rangle\) with
probability 1 by means of a local POVM measurement \(M_i\) on
Clare followed by local unitary operations \(U_{A}(i)\) and
\(U_{B}(i)\) on Alice and Bob, respectively.
\end{theorem}

{\em Proof.  }
We prove that we can always choose a local POVM \(M_i\) on Clare,
local unitary operations \(U_{A}(i)\) and \(U_{B}(i)\) on Alice and Bob
(depending on the outcome \(i\) of the POVM \(M_i\)), such that
\begin{equation}
\label{eq:2epr_psi}
|\Psi\rangle = U_{A} (i) \otimes U_{B} (i) \otimes M_i
(|000\rangle + |011\rangle + |102\rangle + |113\rangle) \quad
\forall i,
\end{equation}
where \(\sum_{i}M_i^\dag M_i =\openone\).
In terms of the "flattened" matrix form \(\tilde{\Psi}\) where the indices
\((i_1,i_2)\) are concatenated, Eq.~(\ref{eq:2epr_psi}) is rewritten as
\begin{equation}
\tilde{\Psi} = [U_A (i) \otimes U_B (i)] \openone M_i^T \quad \forall i.
\end{equation}
By choosing \(M_i^T = (M_i^\ast)^\dag
=(U_A (i) \otimes U_B (i))^\dag \tilde{\Psi}\), it should be
satisfied that
\begin{align}
\openone &= \sum_{i}(M_i^\ast)^\dag M_i^\ast \nonumber\\
&= \sum_{i} [U_A (i) \otimes U_B (i)]^\dag \tilde{\Psi}\tilde{\Psi}^\dag
[U_A (i) \otimes U_B (i)].
\end{align}
Such a local POVM \(M_i\) always exists, because we can depolarize any
\(\tilde{\Psi}\tilde{\Psi}^\dag\) to the identity \(\openone\) by
random local unitary operations \(U_A (i) \otimes U_B (i) \) on Alice
and Bob \cite{bennett+96,mosca+00}.
This randomization can be alternatively achieved by applying a set of 16 local
unitary operations \(\sigma_A^\mu \otimes \sigma_B^\nu\) with equal
probabilities, where \(\sigma^\mu\) and \(\sigma^\nu\) (\(\mu,\nu = 0,1,2,3\))
are the Pauli matrices. This completes the proof.   \qed

\begin{theorem}
Consider \(l\)-partite pure states in the Hilbert space \({\mathcal H} =
\C^{k_1} \otimes \C^{k_2} \otimes \cdots \otimes \C^{k_{l-1}} \otimes
\C^{k_1 \times k_2 \times \cdots \times k_{l-1}}\),  the maximally entangled 
state, which is the \((k_1\times\cdots\times k_{l-1})\times 
(k_1 \times\cdots\times k_{l-1})\)
identity matrix \(\openone\) in concatenating the indices
\((i_1,\ldots,i_{l-1})\), can create any state with probability 1 by means of
a local POVM on the \(l\)-th party followed by local unitary operations
on the rest of the parties.
\end{theorem}

{\em Proof.  }
The generalization of the proof in the \(2\times 2\times n\) case is
straightforward. \qed

These theorems suggest that when one of multiparties holds at least a half
of the total Hilbert space, the situation is somehow analogous to the
bipartite cases.
The maximally entangled state, i.e., the representative of the generic class,
can create any state with certainty.

%
\section{Extension to mixed states}
\label{sec:mixed}
In this section, we extend the onion-like SLOCC classification of pure
states in Sec.~\ref{sec:class} to mixed states.

A multipartite mixed state \(\rho\) can be written as a convex combination
of projectors onto pure states (extremal points),
\begin{equation}
\label{eq:rho_decomp}
\rho = \sum_{i} p_{i}
|\Psi_{i}(\mathcal{O}_{i})\rangle\langle\Psi_{i}(\mathcal{O}_{i})|,
\quad p_{i}>0,
\end{equation}
where each pure state \(|\Psi_{i}(\mathcal{O}_{i})\rangle\) belongs to one of
the SLOCC entangled classes (i.e., an SLOCC orbit \(\mathcal{O}_{i}\)).
Our idea is to discuss, in Eq.~(\ref{eq:rho_decomp}), how \(\rho\) needs
at least an outer entangled class \(\mathcal{O}_{\rm max}\), among the
set \(\{\mathcal{O}_i\}\), in the onion structure of Fig.~\ref{fig:onion}.
That is, we are interested in the minimum of \(\mathcal{O}_{\rm max}\)
for all possible decomposition of \(\rho\).
Because the onion picture is divided by every
{\em SLOCC-invariant closed} subset (i.e., every SLOCC orbit closure) of pure
states, their convex combination in Eq.~(\ref{eq:rho_decomp}) constitutes the
{\em SLOCC-invariant closed} convex subsets of mixed states (see
Fig.~\ref{fig:mixed}.).
Note that, in the onion picture of the multipartite pure cases, there can be
"competitive" closed subsets which never contain nor are contained by
each other.
An example is the closures of three biseparable classes \(B_i\) in
Fig.~\ref{fig:onion}. So, in the extension to mixed states, we should assemble
all subsets of mixed states which require at most these biseparable classes
\(B_i\) into one biseparable convex subset by their convex hull.
(The argument is similar to the classification of 3-qubit mixed states
in Ref.~\cite{acin+01}.)

We find that these entangled classes constitute a totally
ordered structure, seen in Fig.~\ref{fig:mixed}, where noninvertible SLOCC
operations can never upgrade an inner class to its outer classes.
For instance, we see that the closure of $W_3$ class of mixed states
(labeled by \(|000\rangle +|011\rangle +|112\rangle\))
is included in the closure of $GHZ_3$-class (labeled by
\(|000\rangle +\frac{1}{\sqrt{2}}(|011\rangle +|101\rangle) +|112\rangle\)).
This classification reflects a diversity of multipartite pure
entangled states a mixed state \(\rho\) consists of:
the outer the class of \(\rho\) is, the more kinds of resources it contains.
Needless to say, it is very difficult to give the criterion to distinguish
convex subsets, even to distinguish the separable convex subset
(i.e., the separability problem), since we face the trouble evaluating all
possible decompositions in Eq.~(\ref{eq:rho_decomp}) for a given \(\rho\).
Let us however prove that the convex combination of nine classes of pure
states gives rise to convex sets that are not of measure zero, in contrast
with the pure case (cf. Ref.~\cite{acin+01}).
This can easily be established with the help of the following lemma:

\begin{figure}[t]
\begin{center}
\includegraphics[clip]{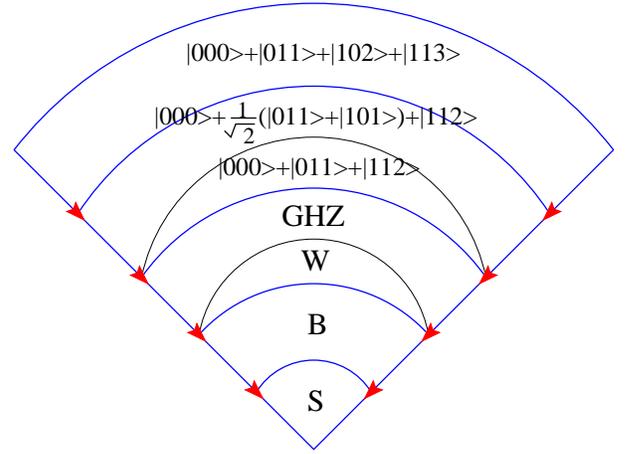}
\caption{The SLOCC classification of multipartite mixed states in the
\(2 \times 2 \times n \;(n \geq 4)\) cases.
Mixed states in the class, labeled by \(|\Psi(\mathcal{O}_{\rm max})
\rangle \in\{ |000\rangle + |011\rangle + |102\rangle + |113\rangle, \ldots,
S \}\), are convex combinations of pure states {\em inside} the "onion skin"
of \(|\Psi(\mathcal{O}_{\rm max})\rangle\) in Fig.~\ref{fig:onion}.
So the outer the class is, the more kinds of multipartite entangled pure
states the mixed states contain.
The edges of the "fan" reflect the structure of extremal points (pure states),
and noninvertible SLOCC operations can never upgrade an inner class to its
outer classes.}
\label{fig:mixed}
\end{center}
\end{figure}

\begin{lemma}
Given two matrices $A,B$ with corresponding ordered singular
values $\{\sigma_i^{A,B}\}$. Denote the ordered singular values of
the matrices $A^TA$ and $B^TB$ as $\{\tau_i^{A,B}\}$. Then the
Hilbert-Schmidt norm \[\|A-B\|_2=\sqrt{{\rm tr}\left((A-B)^\dagger
(A-B)\right)}\] is lower bounded by
\begin{eqnarray*}
\|A-B\|_2&\geq & \sqrt{\sum_i(\sigma_i^A-\sigma_i^B)^2}\\
\|A-B\|_2&\geq & \frac{\|A\|_2}{2(1+\|A\|_2)}\sqrt{\sum_i
(\tau_i^A-\tau_i^B)^2}
\end{eqnarray*}
where we assumed that $\|A\|_2\geq\|B\|_2$.
\end{lemma}
\emph{Proof.} The first inequality can readily be proven using
standard results of linear algebra \cite{horn+85}. The second
inequality can be proven as follows. Defined $X=A-B$; then
\begin{eqnarray}
\|A^TA-B^TB\|&=&\|XA^T+AX^T-X^TX\| \nonumber\\
&\leq& 2\|X\|\|A\|+\|X\|^2
\end{eqnarray}
The left term of this inequality is bounded below by
\begin{equation}
\|A^TA-B^TB\|\geq \sqrt{\sum_i (\tau_i^A-\tau_i^B)^2}.
\end{equation}
The second inequality of the lemma can
now be checked by making use of straightforward algebra.\qed

The fact that a structure of convex sets as depicted in
Fig.~\ref{fig:mixed} is obtained, can now be proven by combining
the previous lemma with the results of the table in
Eq.~(\ref{table}): indeed, it can easily be shown that whenever
there exists a pure state in one class that is separated from all
pure states in another class with a finite non-zero
Hilbert-Schmidt distance, then the corresponding class for mixed
states is absolutely separated from the other one. The previous
lemma guarantees that the Hilbert-Schmidt norm will be non-zero
for all states having a different rank for the matrices $R$ or
$R^TR$ (see the table in Eq.~(\ref{table})). More specifically, all the
W-classes are embedded in the respective GHZ-classes, and the
convex structure as depicted in Fig. \ref{fig:mixed} is obtained.

%
\section{Conclusion}
\label{sec:conclude}
In this paper, (i) we give the complete classification of multipartite
entangled states in the Hilbert space \({\mathcal H}=\C^2 \otimes \C^2
\otimes \C^n\) under stochastic local operations and classical communication
(SLOCC).
Our study can be seen as the first example of the SLOCC classification of
multipartite entanglement where one of multiparties has more than one qubits.
We show that nine classes constitute the five-graded partially ordered
structure of Fig.~\ref{fig:part}. Remarkably, a unique maximally entangled
class lies on its top, in contrast with the \(l\)-qubit (\(l \geq 3\)) cases.
We also present a convenient criterion to distinguish these classes by
SLOCC-invariant entanglement measures.

(ii) We illustrate that important LOCC protocols in quantum information
processing are given as noninvertible (downward) flows between different 
entangled classes in the partially ordered structure of Fig.~\ref{fig:part}.
In particular, we show that two Bell pairs are powerful enough to
create any state with certainty in our situation. Based on these
observations, we suggest that SLOCC classifications can be useful
in looking for new prototypes of novel LOCC protocols.

\begin{acknowledgments}
A.M. would like to express his sincere gratitude to J.I.~Cirac,
M.~Lewenstein, and the members of these groups for stimulating
discussions and warm hospitality during his visit. He also thanks
K. Matsumoto for the discussions on Sec.~\ref{sec:geometry}, as
well as the members of both the University of Tokyo and the ERATO
Project for helpful comments. The work of A.M. is partially
supported by the Grant-in-Aid for JSPS Fellows. The work of F.V. was
supported in part by the E.C. (projects RESQ and QUPRODIS) and the
Kompetenznetzwerk "Quanteninformationsverarbeitung" der
Bayerischen Staatsregierung.

\end{acknowledgments}

%


\begin{thebibliography}{~~}
\bibitem{note_4bit}
Industrious readers might wonder whether there is any relationship between
nine classes in this paper and those in Ref.\cite{verstraete+02_4bit}, which
addressed the SLOCC classification of 4-qubit entangled pure states.
But, this is just an accident, and there is no correspondence.
\bibitem{note_change}
Comparing the number of degrees of freedom in the SLOCC operations with
the dimension of the whole state space, we see whether the number of
SLOCC orbits is finite or not.
The onion picture of Fig.~\ref{fig:onion} suggests how SLOCC orbits merge
each other when the situation changes from the 4 qubits case to
the \(2\times 2 \times 4\) case.
\end{thebibliography}
\end{document}